\documentclass[a4paper,fleqn,usenatbib,useAMS]{mnras}

\usepackage{graphicx}	
\usepackage[flushleft]{threeparttable}
\usepackage{amsmath}	
\usepackage{amssymb}	
\usepackage{multicol}        
\usepackage{bm}		
\usepackage{pdflscape}	
\usepackage{color}
\usepackage{hyperref}
\usepackage{natbib}

\newcommand{\kms}{\,km\,s$^{-1}$} 

\usepackage[T1]{fontenc}
\usepackage{ae,aecompl}

\usepackage{newtxtext,newtxmath}

\title[FVSS-II: PNe kinematics of ICL]{The Fornax Cluster VLT Spectroscopic Survey II - Planetary Nebulae kinematics within 200 kpc of the cluster core}

\author[C.~Spiniello et al.]{C.~Spiniello$^{1,2}$, N.~R.~Napolitano$^{1}$, M.~Arnaboldi$^{2}$, C.~Tortora$^{3}$, L. Coccato$^{2}$,  \and  M. Capaccioli$^{1}$  O. Gerhard$^{4}$, E. Iodice$^{1}$, M. Spavone$^{1}$, M. Cantiello$^{5}$, \and   R. Peletier$^{3}$, M. Paolillo$^{6}$, P.Schipani$^{1}$
\\
$^{1}$INAF - Osservatorio Astronomico di Capodimonte, Salita Moiariello, 16, I-80131 Napoli, Italy \\
$^{2}$European Southern Observatory, Karl-Schwarschild-Str. 2, 85748 Garching, Germany \\
$^{3}$Kapteyn Astronomical Institute, University of Groningen, P.O. Box 800, 9700 AV Groningen, the Netherlands\\
$^{4}$ Max Planck Institute for Extraterrestrial Physics, Karl-Schwarzschild-Str. 1, 85741 Garching, Germany\\
$^{5}$ INAF - Osservatorio Astronomico di Teramo, via Maggini, I-64100, Teramo, Italy.\\
$^{6}$ Dip. di Fisica, Universita\' di Napoli Federico II, C.U. di Monte Sant'Angelo, Via Cintia, 80126 Naples, Italy.
}

\date{Last updated 2018 JFebruary 26}

\pubyear{2018}
\begin{document}
\label{firstpage}
\pagerange{\pageref{firstpage}--\pageref{lastpage}}
\maketitle

\begin{abstract}
We present the largest and most spatially extended planetary nebulae (PNe) catalog ever obtained for the Fornax cluster. 
We measured velocities of 1452 PNe out to 200 kpc in the cluster core 
using a counter-dispersed slitless spectroscopic technique with data from FORS2 on the VLT. 
With such extended spatial coverage, we can study separately the stellar halos of some of the cluster main galaxies and the intracluster light. 
 In this second paper of the Fornax Cluster VLT Spectroscopic Survey (FVSS), we identify and classify the emission-line sources,  describe the method to select PNe and calculate their coordinates and velocities from the dispersed slitless images. 
From the PN 2D velocity map we identify stellar streams that are possibly tracing the gravitational interaction  of NGC~1399 with NGC~1404 and NGC~1387. 
We also present the velocity dispersion profile out to $\sim 200$ kpc radii, which shows signatures of a superposition of the bright central galaxy and the cluster potential, with the latter clearly dominating the regions outside R$\sim 1000$\arcsec($\sim 100$ kpc).
\end{abstract}




\begingroup
\let\clearpage\relax
\endgroup
\newpage

\section{Introduction}
\label{intro}
In the context of hierarchical structure formation and evolution, galaxy clusters represent the final stage 
of the growth of large-scale cosmic structures \citep{Blumenthal84, Davis85,Bahcall88, Navarro96}.  
The Cold-Dark Matter scenario predicts that as the Universe cooled, clumps of dark matter (DM) began 
to condense and within them gas began to collapse, forming the first galaxies which then experienced 
structure growth via the so-called "merging tree" \citep{Kauffmann93, Cole00, DeLucia07}. 
During this growth, various physical processes like tidal interactions, ram pressure stripping, mergers, 
dynamical instabilities, secular evolution and finally gas accretion occurred followed by cooling 
and star formation (e.g. \citealt{Mihos03}). 

All these processes contributed to shape galaxies and the DM surrounding them and they are expected 
to have left their imprints in the outskirts of galaxies and beyond, out to the intracluster regions 
(see e.g., \citealt{Napolitano03, Arnaboldi12}).
Here, dynamical times are longer and galaxy formation mechanisms leave signatures of gravitational 
interactions such as shells and tidal tails in the kinematics of their components or in the chemical composition 
of the stars \citep{Murante04, Murante07, Bullock05, Rudick06, Duc11, Longobardi15a, Longobardi15b, Pulsoni17}. 

Often the light profiles in the external region of central galaxies in group/clusters change slope with respect 
to the inner stellar light, passing from a S\'ersic to an exponential profile \citep{Seigar07, Donzelli11}. 
This is the case for the Fornax cluster and its central galaxy NGC~1399 \citep{Iodice16, Spavone17}.  
NGC~1399 surface brightness profile has strong variations of the radial slope and ellipticity with radius,  
as demonstrated by deep photometry of the central regions of the cluster from VST 
(\citealt{Iodice16}, hereafter FDS-I) and kinematical studies of the cluster core with globular clusters 
(out to 200 kpc, see Pota et al. 2018, in prep., and \citealt{Schuberth10} for earlier results). 
In particular, FDS-I finds that the averaged light profile of NGC~1399 in g band can be fitted 
with two components that contribute in different percentage to the total light at different 
radii: a S\'ersic profile with n=4.5, which dominates for $R<10\arcmin$ and an exponential 
outer component that contributes about 60\% to the total light and extends out to 
large radii (see Fig. 3 and 12 in FDS-I).  
Moreover, surface density maps of GCs, derived from multi-band wide-field photometry, 
reveal the presence of a complex system of structures and substructures connecting 
NGC~1399 and its intergalactic environment \citep{DAmbrusco16, Cantiello17}. 
Photometry alone is however not sufficient to clarify the nature of this exponential 
profile nor to clearly separate the contribution to the light of the extended halo of NGC~1399 
(stars gravitationally bound to the galaxy) from that of the intracluster light 
(ICL, stars that have been tidally stripped from the outer regions of the galaxy, 
mixing over time to form a diffuse light orbiting in the cluster potential, \citealt{Dolag10}). 
Kinematical information is crucial for understanding if the "excess of light" is made of
diffuse halo stars that are still bound to the surrounding galaxies or by stars that 
are freely flying in the cluster potential.  

A first attempt to distinguish dynamically the contribution of the central galaxy from the cluster environment was made by \cite{Napolitano02},  who argued that the velocity dispersion of the outer regions of NGC~1399 had their kinematics 
modified by the interaction with the potential, concluding that, realistically, the kinematics of the PNe bound to the 
galaxy potential only might flatten to $\sigma\sim260$ \kms. 

Measuring kinematics in the outer parts of Early-Type Galaxies (ETGs) remains, however, very challenging 
since the signal-to-noise of the integrated light spectra drops off before the mass profile flattens. 
Long-slit, multi-object or integral field stellar spectroscopy provide measurements only up 
to few effective radii (R$_{\rm eff}$, e.g., \citealt{Cappellari06, Tortora09}); only discrete tracers can push these limits beyond and help us to understand the formation history of the outer regions of elliptical 
galaxies \citep{Hui95,Arnaboldi96,Mendez01, Napolitano02, Romanowsky03, Douglas07, deLorenzi08, deLorenzi09, Coccato08, Coccato09, Napolitano09, Napolitano11, Richtler11,Forbes11, Pota13,Longobardi15a, Hartke17}.  

As a matter of fact, hints of the presence of kinematically distinct components in NGC~1399 have been found by \cite{Schuberth10} 
in a dynamical study of 700 globular clusters (GCs) out to 80 kpc.  
They showed that the red (metal-rich) GCs might trace the spheroidal galaxy component and can be used to constrain the central DM halo, whereas the blue (metal-poor) GCs show evidence for kinematical substructures from accretion episodes during the assembly of the Fornax cluster. 
These results have been recently confirmed and expanded to larger radii in the companion paper of the Fornax Cluster VLT Spectroscopic Survey (FVSS), namely FVSS-I, by Pota et al. (2018, to be submitted), where GCs velocity are measured out to 200 kpc. 

However, globular clusters usually do not follow the same spatial distribution of the stars and they show most of the time a bi-modal color distribution \citep{Harris91, Ashman98, Brodie06}. 
Planetary Nebulae (PNe) instead provide kinematics that are most of the time directly linked 
to integrated light measurements in ETGs \citep{ Douglas07,Coccato09}. 
PNe represent part of the post-main-sequence evolution of most stars with masses in the range 0.8-8 M$_{\odot}$, which means that they are drawn from the same old/intermediate population that composes most of the light in ETGs \footnote{We note that recent studies (e.g. \citealt{Bertolami16}) point to the fact that the progenitors of PNe could be older than previously expected, up to 9-10 Gyrs (especially for the bright ones)}. 
PNe are sufficiently bright to be detected also in external galaxies in the local universe, 
and they are easier to detect at large galactocentric radius where the background continuum is fainter. 
Thus, they represent a unique tool to measure the kinematics of elusive stars 
in low surface brightness regions where they are easily observable 
through their [OIII] emission at 5007 \AA. 
Measuring kinematics of Intracluster PNe (ICPNe) is a very effective way to measure the dynamical stage of the intracluster stellar population, 
even commonly called intracluster light (ICL), and to assess how 
and when its light originated \citep{Napolitano03, Murante04, Murante07, Gerhard07}

Traditionally, PNe have been detected using an on-band/off-band technique \citep{Ciardullo89,Arnaboldi98} where two images of the same portion of the sky are taken, one with a narrow-band filter centered at the redshifted [OIII] $\lambda$ 5007 line and a second with a broad-band filter. Sources in OIII will be then detected in the on-band image and too faint to be detected in the second one, whereas foreground stars will appear in both images with similar brightness. With this technique, however, a spectroscopic follow-up is then necessary to measure PN velocities \citep{ Mendez01,Arnaboldi04, Teodorescu05, Doherty09}.

In parallel, a number of techniques have been developed that allow detection and velocity measurements to be taken in a single step.  One of the most successful is the so-called "counter-dispersed imaging" (CDI), first developed by \cite{Douglas99}. With this technique two images are obtained using a slitless spectrograph with a dispersive element rotated by 180 degree between the two exposures. 
In this way, the emission lines objects appear as unresolved dots in each dispersed image while stars appear as elongated streaks in the direction of dispersion with a length determined by 
the spectral resolution and the filter full width half maximum (FWHM). 
Due to the fact that the at the two position angles (PAs), the grating disperses the monochromatic emissions in two opposite directions, the PNe will be shifted in the two rotated images by an amount proportional to their velocities. 
By registering the two images using dispersed spectra of foreground stars, matching the pairs of unresolved emitters and measuring the distance between them, one can simultaneously identify PNe and measure their velocities.

The use of CDI was demonstrated to be so successful and efficient that \cite{Douglas02} decided to build a dedicated slitess spectrograph to study the kinematics of extragalactic PNe: the Planetary Nebuale Spectrgraph (PN.S) mounted at the 4.2m William Herschel Telescope.  
In the last 15 years the PN.S has produced a conspicuous number of referred publications and has substantially contributed to measure kinematics of galaxy halos \citep{Romanowsky03, Merrett06, Douglas07,Noordermeer08, Coccato09, Napolitano09, Cortesi13}, and of galaxy disk \citep{Aniyan18}. 
In order to extend the study of diffuse outer halos of ETGs using PNe as tracers for galaxies situated in the Southern Hemisphere, we obtained counter dispersed images with the FORS2 at the VLT. To carry out a complete mapping of an area of $50\arcmin \times 30\arcmin$ centered in the core of the Fornax cluster, for which we have already deep imaging coverage with VST (ugri) and VISTA (J and K), we obtained a total of twenty FORS2 pointings each covering an area of $\sim 6.8\arcmin \times 6.8\arcmin$. 
Complementing this data with previous CDI FORS1 observations of $\sim 180$ PNe around NCG~1399 and NCG~1404 published in \cite{McNeil10}, hereafter MN10,  we are able to obtain velocity and velocity dispersion profiles up to $\sim30\arcmin$ from the central galaxy NCG~1399 ($\sim6$ times the effective radius, R$_{\rm eff} \sim 5$ arcmin, as reported in FDS-I).

With the spectroscopic maps of the central 200 kpc of the Fornax cluster, PNe (and GCs) can be traced out to intracluster regions where they allow us to probe the cluster potential together with other satellite systems like ultra compact dwarfs (UCDs) or dwarf galaxies (see e.g. \citealt{Prada03}). 
The simultaneous constraints from independent tracers will allow us to study in great detail the transition region where the cluster potential starts to dominate the 
galaxy sub-haloes. Here one can expect to isolate unmixed substructures in the phase space, as the relics of 
interactions of satellite disruptions \citep{Napolitano03, Arnaboldi04,Bullock05, Arnaboldi12, McNeil10, Romanowsky12, Coccato13, Longobardi15a}.

In this paper we present positions and velocities of 1635 Planetary Nebulae (1452 of which are new detections), 
describe the data reduction and calibration (Section 2), explain our techniques and methods to identify planetary nebulae (Section 3) and infer their velocities directly from the calibrated and registered images (Section 4). 
A more complete and detailed description of the techniques is given in MN10, we therefore refer the reader to that paper for detailed information on the CDI calibration and data reduction.
In Section 5 we present the final catalog and the histogram of velocities, as well as histograms of velocities of the PNe associated with the three main galaxies and the line-of-sight (LOS) velocity and velocity dispersion distributions as function of radial distance from NGC~1399. 
We extend by a factor of $\sim 8$  the number of detected PNe and by a factor 
of $\sim 4$  the radial spatial coverage of the previous PNe sample. 
Finally, conclusions  and future perspectives are presented in Section 6. 
Throughout the paper, we assume a distance to the core of the Fornax Cluster of 
$20.9\pm0.9$ Mpc from \cite{Blakeslee09}, and therefore $1\arcsec$ corresponds to $\sim$100 pc.
\section{Observations and Data Reduction}
\subsection{Instrumental set-up and calibration}
\begin{figure}
 \includegraphics[width=\columnwidth]{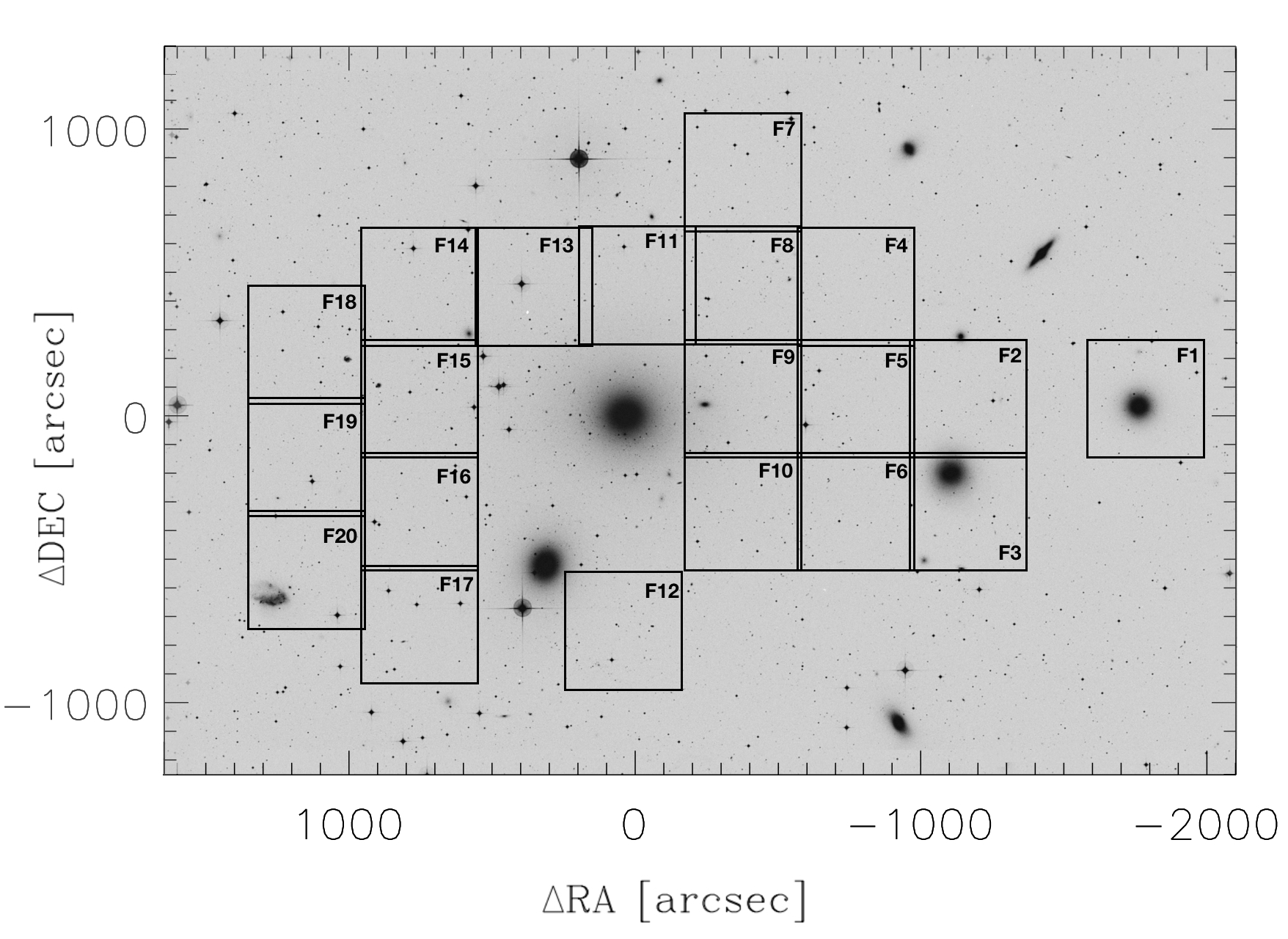}
 \caption{DSS Image of the Fornax Cluster Core ($60\arcmin \times 40\arcmin$) centered around NGC~1399. Black numbered boxes show the 20 FORS2 pointings obtained in P96. }
 \label{fig:FORS_pointings}
\end{figure}

The observations have been acquired in P96 (096.B-0412(A), PI: M.~Capaccioli) from November 2015 to  December 2016. 
Twenty pointings, for a total of 50 hours of observing time were carried out with FORS2 on the 8-mt large ESO Very Large Telescope (VLT), with seeing generally below $0.8\arcsec$ and always below $1\arcsec$, covering a total final area of $\sim 50\arcmin \times 33\arcmin$, centered around $\alpha =$3:37:51.8 and $\delta =-$35:26:13.6. We show the final coverage of the FORS2 pointings (numbered from 1 to 20) on top of a Digital Sky Survey (DSS) image\footnote{The image has been taken from the IRSA at http://irsa.ipac.caltech.edu/data/DSS/.} centered on NGC~1399 in Figure \ref{fig:FORS_pointings}. 

For each pointing, we acquired three scientific exposures, for a total of $\sim 2050$ seconds on target, on each position angle (labeled respectively W0 and W180, were "W" stands for West and the numbers refer to the angle aligned to this direction). The final spatial coverage 
complements\footnote{We did not re-observe the central region already presented in MN10 and we only have a minimal overlap 
between the pointings, necessary to calibrate our PN velocities, as explained in Sec. \ref{velcalib}} the region already observed by MN10 and extends from the center of NGC~1399 out to  $\sim 1000\arcsec$  along the North-South direction and out to $\sim 1300\arcsec$ along East-West, plus a separate pointing centered on NGC1379 ($\sim 1800\arcsec$  west of NGC1~399).  
We used the 1400V grating with a mean dispersion of 0.64 \AA\, pix$^{-1}$ and the OIII/3000+51 filter, centered on the redshifted [OIII] 5007 line (with the SR collimator, $\lambda_{\rm central} = 5054$ \AA, FWHM$=59$ \AA). 
In the focal plane, the presence of the grism causes an anamorphic distortion resulting in a contraction in the direction of dispersion which must be corrected before PN velocities can be calculated (see below). 
With this instrumental configuration, the direction of dispersion is along the row axis (x-axis) but because 
of the slitless technique, both axes retain spatial information. 

We perform all the standard calibration tasks with \textsc{IRAF} and \textsc{IDL}. 
The scientific frames are bias subtracted and flat fielded individually. Cosmic-ray are removed using the routine by \cite{vanDokkum01}. The three scientific exposures of each pointing 
(for the same dispersion direction) are then combined using the \textsc{IRAF} task {\sl imcombine}. 
Finally, the background is computed by smoothing with {\sl fmedian} (30 pixels) each combined frame 
and then subtracting the smoothed image from the original one.

\begin{figure*}
 \includegraphics[width=18cm]{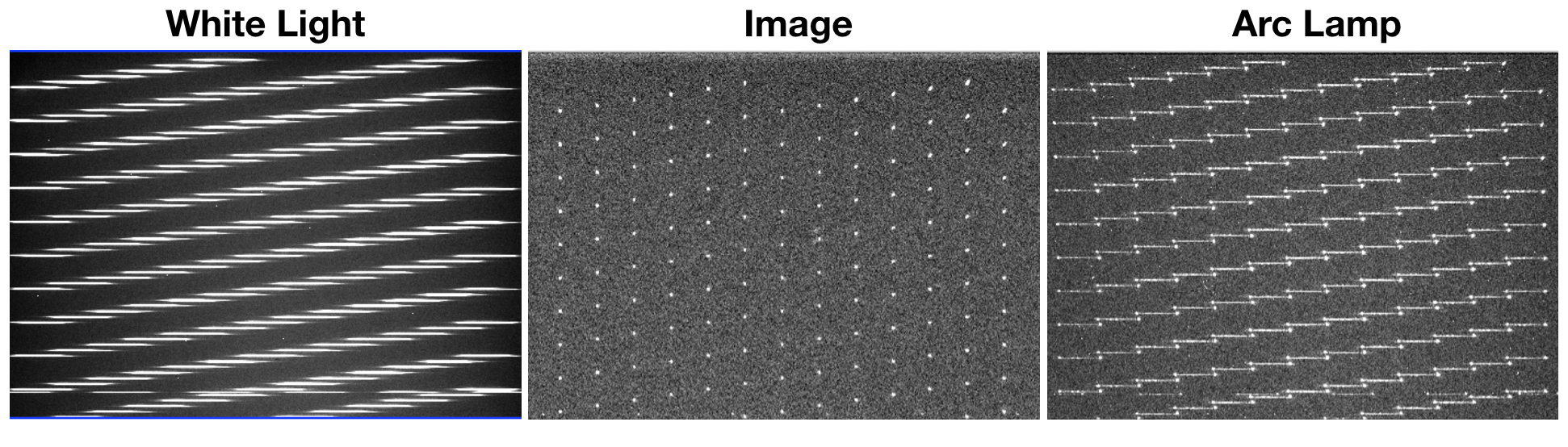}
 \caption{Calibration MXU frames. The left panel shows the white light lamp used to calculate the local dispersion and to measure the bandpass shift. Middle and right panels show the image and the arc lamp used to map the anamorphic distortion. We use the standard HgCdHeAr arcs template and, in particular, two He lines (4921.93\AA, and 5015.67\AA) are transmitted through the narrow filter bandpass and are visible in the rightmost panel.}
 \label{fig:Calib}
\end{figure*}

In addition to this, we also require special calibrations to take into account
the effect of the dispersing optical element on our images. We use a MXU mask with a uniform array of slits. 
We requested to illuminate the mask with white light,  then we requested to add a grism to the light-path 
and finally to disperse an arc lamp through the same filter ([OIII]3000+51).  
These three calibration frames, showed in Figure~\ref{fig:Calib}, allow us to solve for the local dispersion, the bandpass filter shift, and the anamorphic distortion introduced by the grism.

In order to calibrate and correct our images, we follow the recipes of MN10, using the same \textsc{IRAF} and \textsc{Mathematica} tasks and complementing them with self-written \textsc{IDL} scripts.  We refer the reader to that paper for an extensive data-reduction overview.  
Here we briefly list the three main corrections we applied and the procedures we used to register the images, which is necessary to identify the PNe candidates.
\begin{itemize}
\item {\sl Mapping the anamorphic distortion}: 
we use the x,y-positions from the image of the slitlets and the positions of the brightest line ($\lambda5015.67$)  in the dispersed arc lamp calibration frame to map the anamorphic distortion introduced by the grism.  We tabulate the difference in x and y from the slit x$_{\rm s}$ and y$_{\rm s}$ to the monochromatic x$_{\lambda_{1}}$, y$_{\lambda_{1}}$, and use it to make a map of the anamorphic distortion with the \textsc{IRAF} task {\sl geomap}. Using {\sl geotran}, we then applied the solution from {\sl geomap} to all of our images to remove such a distortion.
\item {\sl Calculating the local dispersion}: 
we measure the distance between two lines in the dispersed arc lamp.  
Assuming that the dispersion is linear on this scale, we then convert our dispersions from the measured spectral-plane dispersions to image-plane dispersions using \textsc{IRAF} tasks and \textsc{IDL} routines. 
\item {\sl Measuring the bandpass shift}: we map the bandpass filter shift as a function of position. We
do this by comparing the Gaussian center of the dispersed white light to the position of the $\lambda5015.67$ line. As in MN10 we find the underlying smooth 2-D function 
that describes the bandpass shift caused by the presence of the interference filter into a converging beam and we model it using \textsc{Mathematica}.
\item {\sl Registration of the images:}
Before calculating velocities, we must register the W0 and W180 corrected and calibrated frames in order to have 
all the strips and extended objects in the same relative positions in each pair. 
The registration is done by applying a rigid shift calculated by mapping the position of few bright stars, 
after adjusting them for the bandpass shift (see MN10). 
The location of a star in our images is always measured with a 2-D Gaussian fit using the {\sl n2gaussfit} task in \textsc{IRAF}.  
For each pointing, we measured the positions of $\sim 15$ profiles of stars spanning 
the whole field-of-view of the pointing in the spectral plane, we apply the described bandpass shift, correct for anamorphic distortion and rotate the corrected positions of the W180 frame. 
\end{itemize}
Spectral and image planes are defined in \cite{Arnaboldi07}: the spectral plane is the dispersed image as observed, and the image plane has been corrected for the distortion introduced by the grism thereby making an undispersed, distortion-corrected image for one wavelength. 
In general, to switch from the spectral-plane, where the bandpass shift is measured, to the image-plane, that we need to use to identify PNe, we use the \textsc{IRAF} task {\sl geoxytran}.

\begin{figure*}
 \includegraphics[width=18cm]{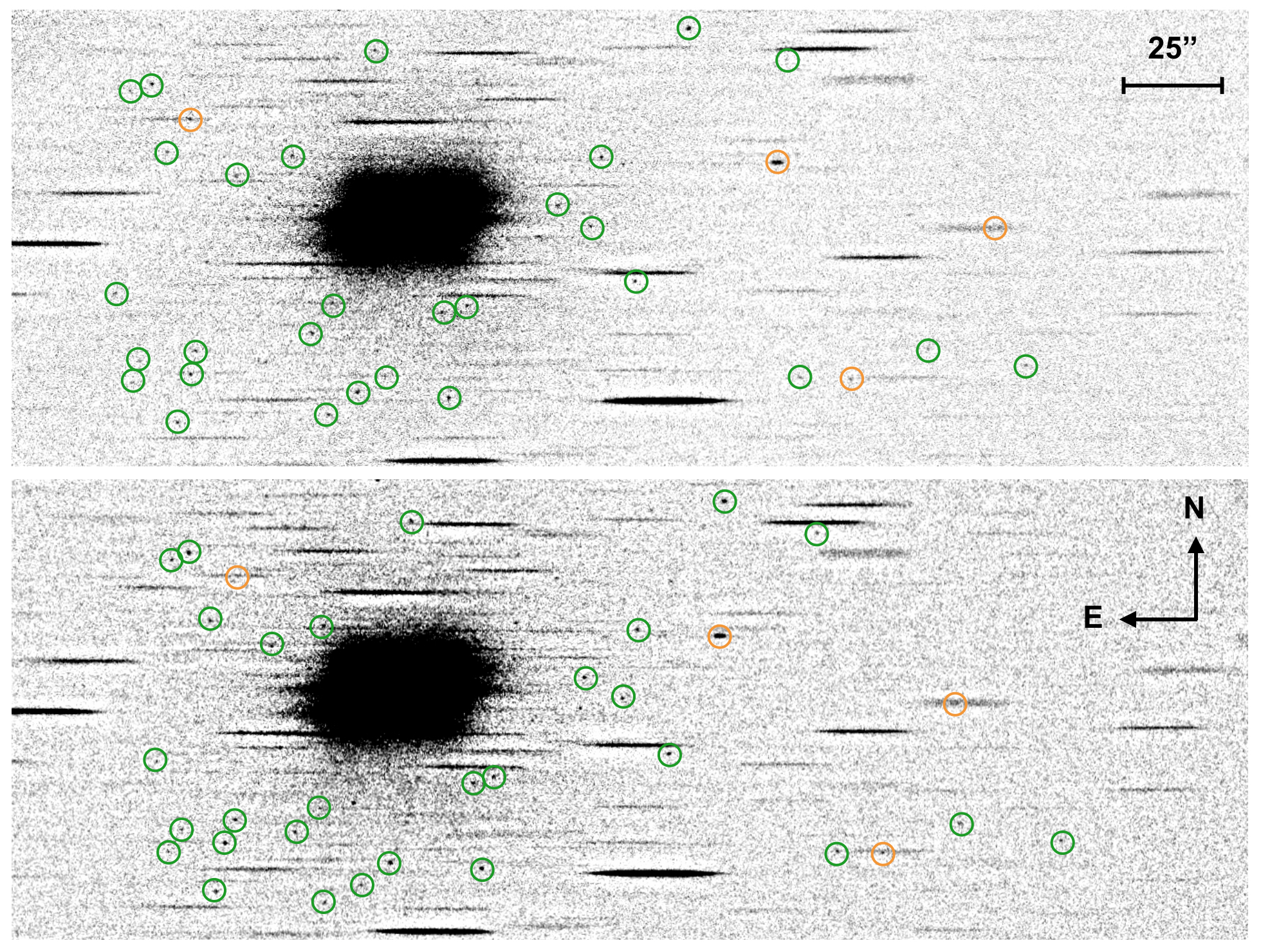}
 \caption{Part of a single pointing (Field 3). The {\sl top frame} is the W0 exposure while the {\sl bottom frame} is the counter-dispersed W180 exposure rotated back to be compared to the first. The big diffuse light on the left center of each frame is NGC~1387 while the other horizontal strikes are stars. The two images have been corrected, registered and calibrated (see text for more details) in a way that the galaxy and the stars appear in the same position. Only planetary nebulae and point-like emitters have different x-position in the two panels. However, the [OII] doublet (the emission from galaxies at z $\sim 0.347$) is resolved thanks to the adopted grism and most of the background Lyman-$\alpha$ galaxies are associated with an elongated continuum. Some examples (not all of them, for clarity of the image) of emitters are circled (PNe in green, background objects in orange).}
 \label{fig:PNe_search}
\end{figure*}

\section{Candidates Identification}
The CDI is a slitless technique that makes use of two counter dispersed frames (taken by rotating the position angle of the field by 180 deg) where the light is selected by a [OIII] filter: here Oxygen emission-line objects (like PNe) appear as spatially un-resolved monocromatic sources, while continuum sources (stars, background galaxies) show up as strikes or star-trails. The velocity of the PNe is obtained by measuring the displacement of the [OIII] emission with respect to a calibrated frame (see MN10 for further details). 
Thus, CDI allows both the detection and measurement of the Doppler shift with a single observation. 
Figure~\ref{fig:PNe_search} shows an example of the best-quality single FORS2 field (FIELD3) taken in the two position angles, W0 and W180. 
In these images, that have already been corrected, calibrated and registered, monochromatic spatially unresolved objects are easy to spot because they appear as unresolved sources. 
Emitters will appear in the two images at the same y-position,  with similar intensity (not necessarily identical, given that the images quality might be slightly different) but shifted in x-direction of an amount which is indeed proportional to their line-of-sight (LOS) velocity.

Standard criteria to identify PNe have been established and successfully tested by the PN.S community and previous authors \citep{Arnaboldi02, Romanowsky03, Douglas07, McNeil10, McNeil12, Pulsoni17}. In particular, they are based on the identification of emission-line objects that are then classified into three main classes:
\begin{itemize}
\item[1)] {\sl PNe candidates}: they are monochromatic sources spatially unresolved in both wavelength and space. 
\item[2)] {\sl  Lyman-$\alpha$ galaxies}: these are Lyman-$\alpha$ galaxies at z $\sim3$ that are most of the time associated with a continuum and therefore can be easily excluded from our final PNe catalog (see Fig.~\ref{fig:PNe_search}). To estimate the contamination of Lyman-$\alpha$ emitters not associated with a continuum we use a very simple argument: there is no reason to believe that the number of background galaxies changes for different velocity bins. Directly from the final histogram of velocity that we present in the next section, we see that we find only few objects (4) in the velocity range $2500$ -- $3000$ \kms. This implies that on the full velocity range ($0$--$3000$), we expect to find $\lesssim 30$ Lyman-$\alpha$ contaminators, which represent $\lesssim 2$\% of our total sample. \item[3)]{\sl [OII] emitters at z $\sim 0.347$}: at this redshift the [OII] doublet will fall in our narrow band filter and therefore objects will appear bright in the images. However, thanks to spectral resolution of the grism (GRIS\_1400V), 
we are able to resolve the doublet (see Fig.~\ref{fig:PNe_search}) ensuring in this way a minimal amount of contamination from this class of objects in the final PNe catalog.
\end{itemize}

In Figure~\ref{fig:PNe_search} we highlight with green circles our PNe candidates and with orange ones the other background emission-line objects. 

For the identification of the PNe we rely on visual inspection: we blink the two counter-rotated, registered and shifted images and select point-like sources with same y-position and different x-positions and not associated with a continuum. 
Moreover, to identify PNe at the very center of galaxies, where the integrated stellar light dominates, 
we subtract the W180 image from the W0 frame and we search for emission-line objects as positive/negative residuals.

In order to make the procedure as objective as possible and to limit the spurious detections, sources are identified by two different members of our team, that have checked the registered images in a completely independent way.  
A PN is added to the final catalog only if it is confirmed by both researchers independently.

\section{Planetary Nebulae line-of-sight Velocities}
As already discussed, the counter-dispersion shifts the x-position of the PNe that appear in different places in the W0 and the W180 images. The  line-of-sight (LOS) velocity of a PN is thus a function of the separation between the two positions: 

\begin{equation}
 \lambda=\lambda_{0} + \dfrac{d\lambda}{dx} \dfrac{\Delta x}{2}
\end{equation}
where $\lambda$ is the measured wavelength of the planetary, $\lambda_{0}$ is the central wavelength of the passband narrow filter (used in the registration procedure), $d \lambda/dx$ is the local dispersion and finally $\Delta x$ is the separation, in pixels, between the position of the same PN in the W0 and W180 frames.

\subsection{Errors on the velocity measurements}
The uncertainties on velocity measurements of single planetary nebulae come from three difference sources: i) the uncertainties on the PN position in each image (W0 and W180) -  roughty corresponding to half of a pixel, 
 ii) the uncertainties on transformation from spectral-plane to image-plane necessary to correct for anamorphic distortion and local dispersion. iii) The uncertainties on the bandpass shift. 
We used the 1400V  grating with a measured mean dispersion of 0.6396 \AA px$^{-1}$. 
Thus, propagating these errors from the formula in Eq.~1, we obtain uncertainties on the single velocity measurements. 
These are of the order of 30-45\kms, perfectly consistent with the errors given in MN10 and in \cite{Mendez01} and slightly larger than the $20$\kms errors in the Planetary Nebula Spectrograph data of \cite{Douglas02}.
Detailed information on the errors on each PN velocity will be provided with the full catalog (positions, v$_{\rm LOS}$ and m$_{5007}$) in a publication in preparation. 

\subsection{Absolute velocity calibration}
\label{velcalib}
\begin{figure}
 \includegraphics[width=\columnwidth]{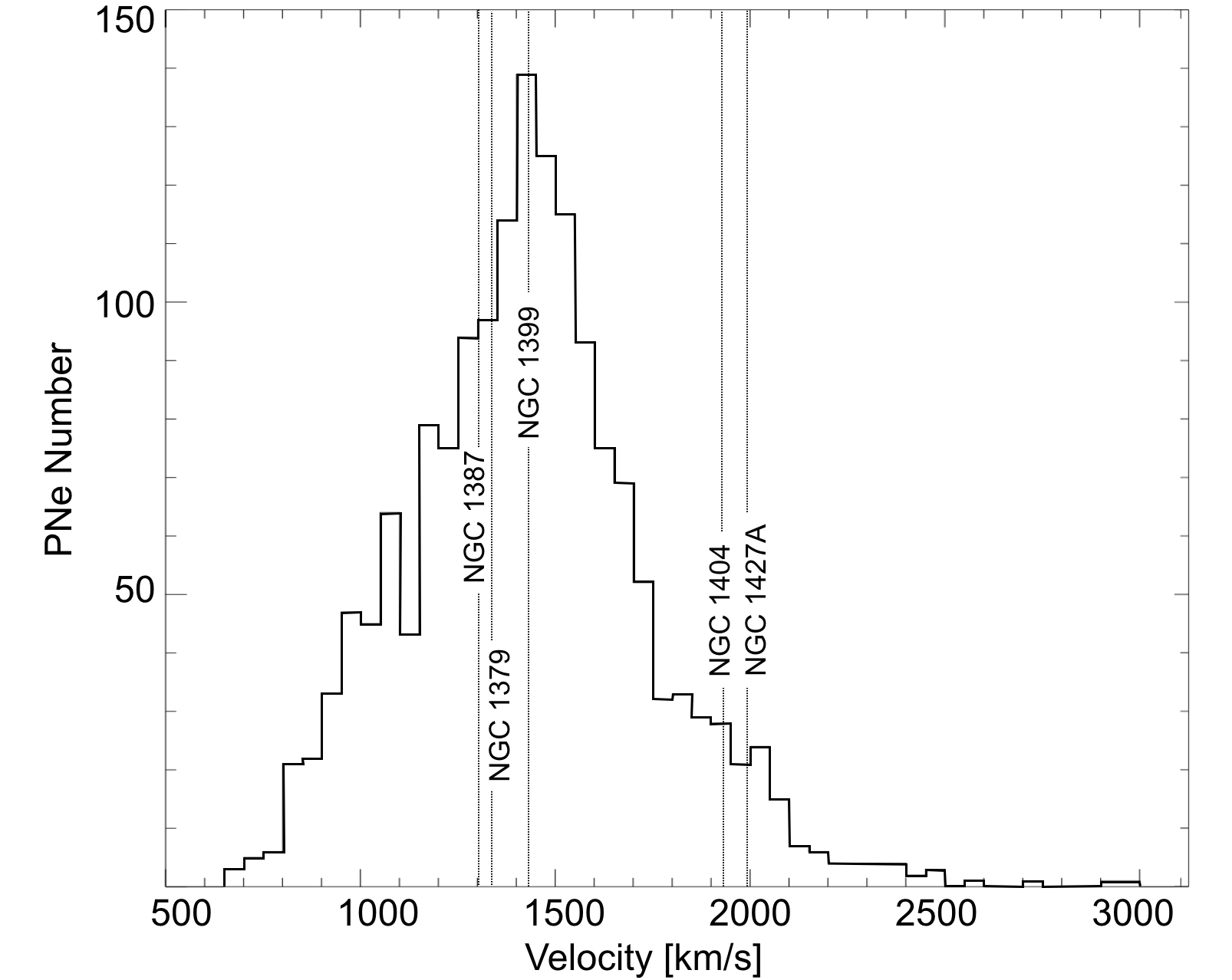}
 \caption{Histogram of velocities of the full catalog of $1635$ PNe, including MN10 data. 
 Vertical dotted lines show the systemic velocities of the Fornax Galaxy members observed within our pointings. }
 \label{fig:velhist}
\end{figure}

The LOS PN velocities cause a shift of the monochromatic emission with respect to the LOS velocity correspondent to the central wavelength of the narrow band filter. 
The mean velocity of the full catalog is $1433$\kms with a standard deviation of $312$\kms, calculated after applying heliocentric correction to each field separately. 
In Figure~\ref{fig:velhist} we show the total, global histogram of velocities. 
In the plot, we also include the PNe presented in MN10 to have a homogeneous spatial coverage from NGC~1399 outwards. Vertical arrows highlight the systemic velocities of the Fornax galaxy members observed within our pointings.  We cover NGC~1379 and NGC~1387. We also complete the coverage on NGC~1404 already partially observed by MN10 and finally, we also observe the irregular galaxy NGC~1427A. \footnote{ Rather than being disrupted during its first passage through the cluster as reported in \citealt{Chaname00}, a recent paper by \citealt{Lee-Waddell18} claims that the irregular optical appearance of NGC~1427A might have tidal origins.}

We calibrate our line-of-sign velocities (LOSV) measurements on the basis of those by MN10 by comparing the 14 PNe in common between the two samples.  
MN10 objects were in turn calibrated against the 1994 NTT multi-object-spectroscopy measurements of 
\cite{Arnaboldi94}\footnote{Since our field pointings strategy is complementary to the observations of MN10, who covered the central region of NGC~1399, we do not have objects in common with \cite{Arnaboldi94}, that are restricted to the innermost 200 arcsec from NGC~1399. This is the reason why the two-steps calibration is adopted.}. 
MN10 found a systematic offset of +166 \kms\, with respect to \cite{Arnaboldi94}, which was independent of the position on the CCD. For the few (14) PNe in common between us and MN10, we find a negative systematic offset of 46\kms\, between our velocities and those of MN10, that we apply as zero velocity offset to our catalog.  
The objects in common with MN10 are found in three different fields, pointing at three different sky positions.  Moreover, we detect few (3) objects in an overlapping region between two of our different fields. For them we obtain two independent measurements of velocity, that resulted to be in perfect agreement. 
We are therefore confident that the offset does not change from pointing to pointing. 

\begin{figure}
 \includegraphics[width=\columnwidth]{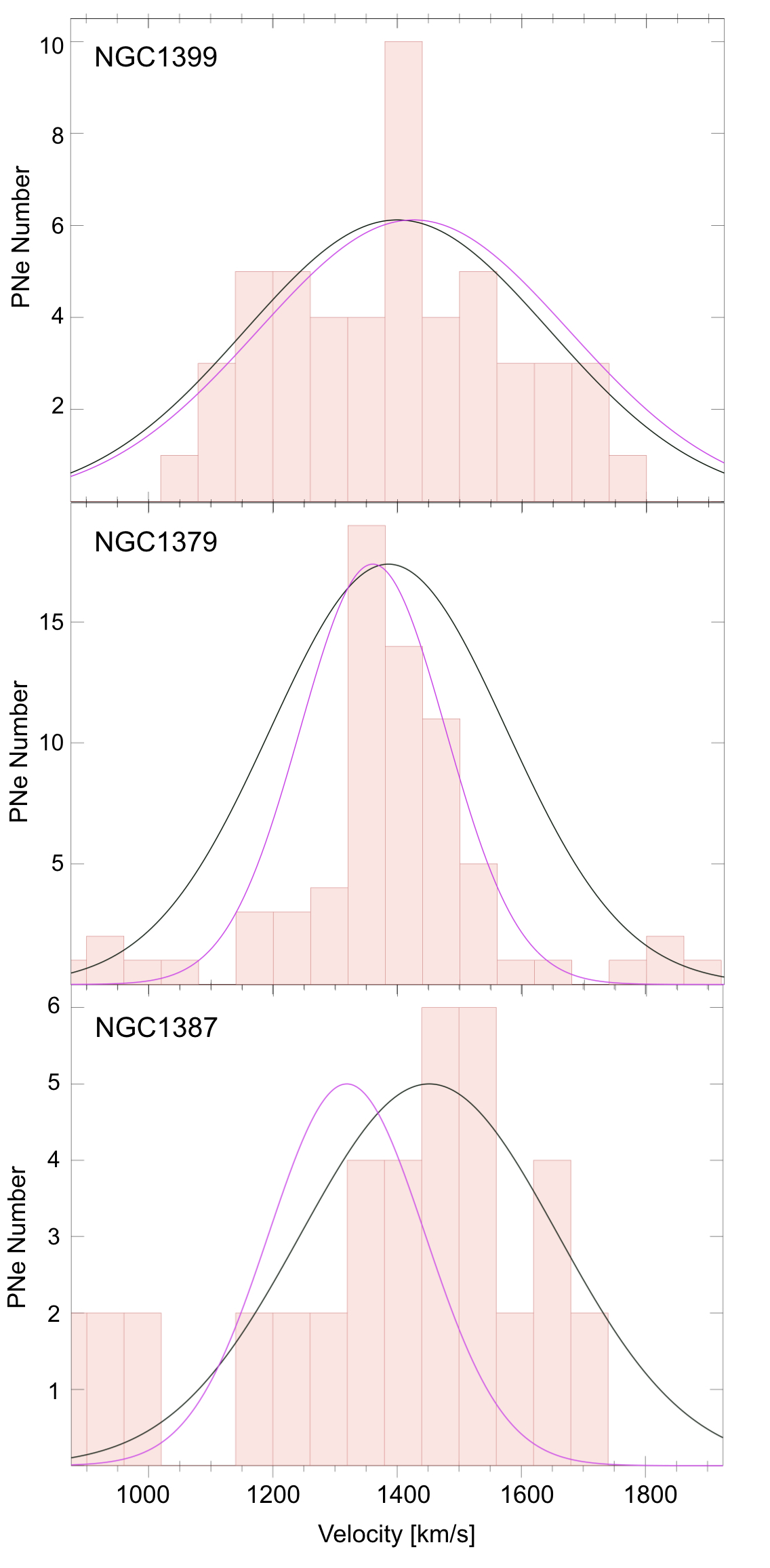}
 \caption{Histograms of velocities for the PNe within the effective radius of NGC~1399 (upper panel, from MN10), NGC~1379 (in  FIELD 1) and NGC~1387 
 (between FIELD 2 and 3). In each panel, the black gaussian line is centered on the median velocity inferred by the PNe distribution with its standard deviation, and the magenta gaussian line is centered on the systemic velocity and has a standard deviation equal to the tabulated central velocity dispersion of the galaxy, corrected for aperture, using the formula of Cappellari et al. (2006).  An overall good agreement is found, although the median velocity calculated for NGC~1387 is $\sim 100$\kms larger than the one inferred from a spectrum centered on core of the galaxy (see the text for a possible explanation for this partial disagreement).}
 \label{fig:sys_vel}
\end{figure}

\begin{figure*}
 \includegraphics[width=18cm]{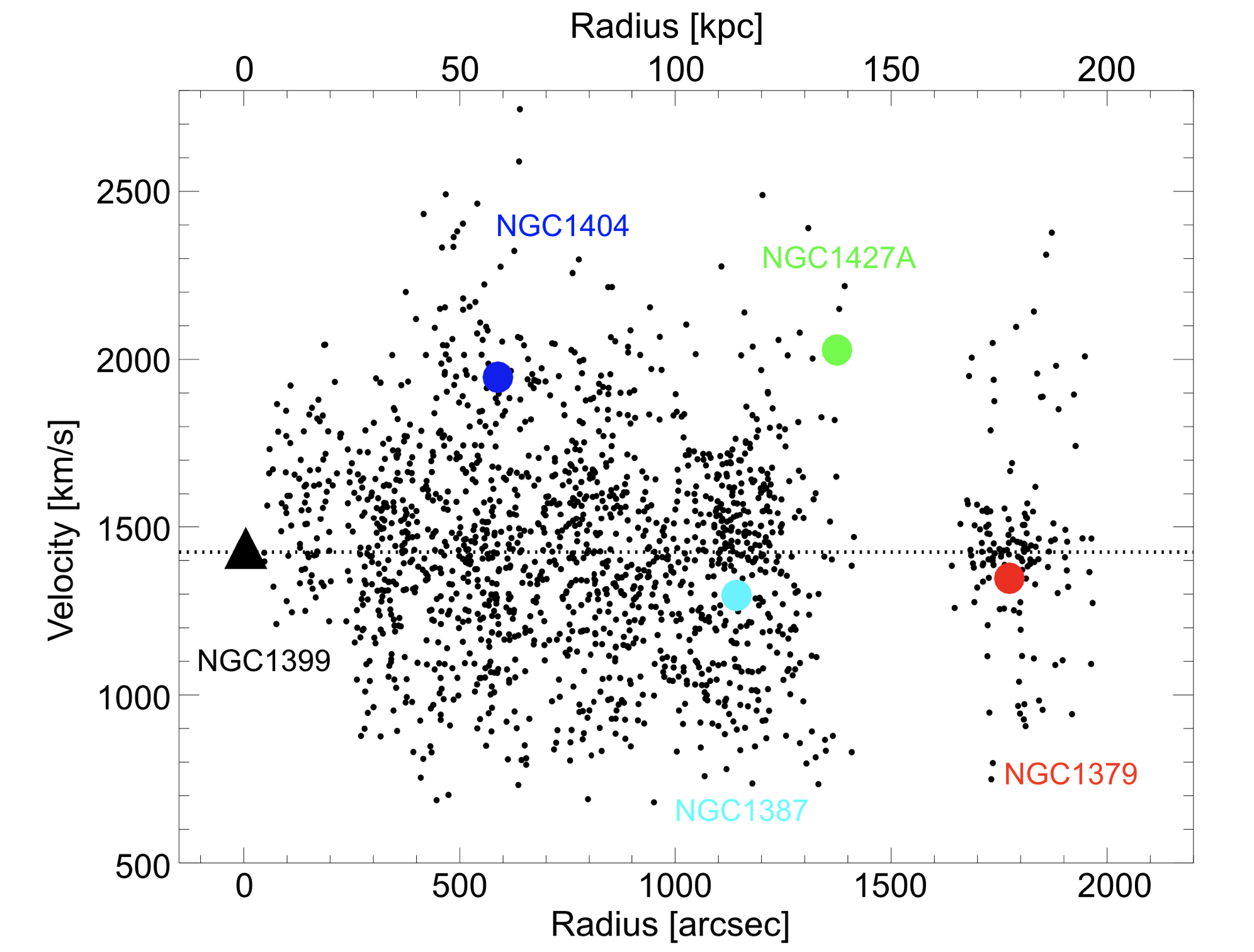}
 \caption{Phase space distribution diagram. NGC~1399 is plotted as black triangle and its systemic velocity is shown as dotted black line. The colored circles represent the other cluster members, with names plotted in the figure with the same color as the symbol. Planetary nebula velocities are shown as a function of radius from NGC1399 in black (black small circles).}
 \label{fig:phasespace}
\end{figure*}
  
To further confirm our velocity calibration, we focus on fields where we observe galaxy members of the Fornax cluster: NGC~1379, with a systemic velocity of V$_{\rm sys} = 1324$\kms\, and an effective radius of R$_{\rm eff} = 23.3\arcsec$ \citep{Caon94}, observed in FIELD1; 
NGC~1387, with a systemic velocity of V$_{\rm sys} = 1302$\kms\, and R$_{\rm eff} = 42\arcsec$ \citep{deVaucouleurs91}, observed in FIELD2 and FIELD3 (see Fig.~\ref{fig:FORS_pointings}) and 
finally the central galaxy NGC~1399 (V$_{\rm sys} = 1425$\kms\, and R$_{\rm eff} = 303\arcsec$, FDS-I\footnote{We note that FDS-I recently recomputed the effective radius for NGC~1399, originally given in \citealt{Caon94}, using deep photometry in g- and i-band and obtaining larger values of R$_{\rm eff, g}=5.87 \pm 0.10$ arcmin and R$_{\rm eff, i}=5.05 \pm 0.12$ arcmin}), mainly covered by MN10. 
For NGC~1399 and NGC~1379, systemic velocities were taken from the NASA/IPAC Extragalactic Database (NED\footnote{https://ned.ipac.caltech.edu}). For NGC~1387 we calculate the velocity directly from a spectrum obtained with The Wide Field Spectrograph (WiFeS) at the Australian National University 2.3 Telescope\footnote{The spectra have been kindly shared with us by Kenneth Freeman and Mike Bessell, to whom we are grateful. }. 
We take all the PNe within one effective radius of each galaxy, assuming that the probability that they 
are bound to such galaxy is high, and calculate their median velocity, after applying the absolute velocity calibration described above. 
Figure~\ref{fig:sys_vel} shows the histograms of velocities for the PNe associated to these three galaxies. 
In each panel, the black Gaussian line is centered on the median velocity inferred by the PNe distribution and broadened by the standard deviation of the PN velocities, and the magenta gaussian line is centered on the systemic velocity from the NED Database and has a standard deviation equal to the tabulated central velocity dispersion of the galaxy, further aperture corrected to the R$_{\rm eff}$ using the formula in the Eq.1 of \cite{Cappellari06}. 
We caution the reader on the fact that not all the PNe spatially close to a given galaxy must in fact be bound to it. This could cause the overestimation of the standard deviation of the PNe, which is larger than the stellar velocity dispersion. A more rigorous bounding criterion will be defined in a forthcoming paper of the current series. 
For two out of the three systems, we found a very good agreement. Only for NGC~1387, we infer a velocity which is $\sim 100$ \kms larger than that tabulated in the literature. However, we note that the two measurements are within 1$\sigma$ of each other. 
We believe that part of this difference could be caused by the different image quality between FIELD2 and FIELD3 combined with the fact that this galaxy (and the PNe bound to it) shows clear sign of rotation.   
In fact, FIELD3 points to the receding part of the galaxy and is the field with the best image quality we have (in this pointing we found a total of 143 PNe). On the contrary, FIELD2, on the approaching side of the galaxy rotation, has lower image quality (in total 95 PNe). Consequently, the number of PNe with receding velocity with respect to the systemic velocity of NGC~1387 is larger than the number of PNe with approaching velocity. 
We finally note that the PNe number within $1$ R$_{\rm eff}$ of NGC~1387 is lower than the number found for the other two galaxies (most probably because NGC~1387 might have a lower PN specific density and a smaller effective radius\footnote{For a definition of the PN luminosity-specific number and its variation as function of galaxy type and colour we refer to \citep{Buzzoni06}}), making the statistics in the histogram poorer. Finally, we will show in the following sections that hints for gravitational interaction between NGC~1399 and NGC~1387 are  suggested by the presence of a stream of high-velocity PNe connecting the two galaxies. Some of these high-velocity PNe might 
overlap spatially with the PNe gravitationally bound to NGC~1387, thus broadening the PN LOSVD at the location of this galaxy.

It is worth noticing that this kinematical stream is in the same spatial region where FDS-I found a $\sim 5$ arcmin long faint stellar bridge ($\mu_{g}\sim29-30$ mag arcsec$^{-2}$) and \cite{DAmbrusco16} and \cite{Cantiello17} reported an over density of the distribution of blue GCs. We argue that the stream is the result of the stripping of the outer envelope of NGC~1387 on its east side. We will further speculate on this point in the following section.

We thus conclude about the robustness of our velocity calibration. In fact, we reproduce convincingly the systemic velocities of the three galaxies that fall within our pointings and we find a good agreement (after the shifting) with the results of MN10. 
Further confirmation can be also found in Figure~\ref{fig:velhist} where we show the final histogram of velocities and overplot the  systemic velocities of the observed Fornax galaxy members. 

\begin{figure*}
 \includegraphics[width=18cm]{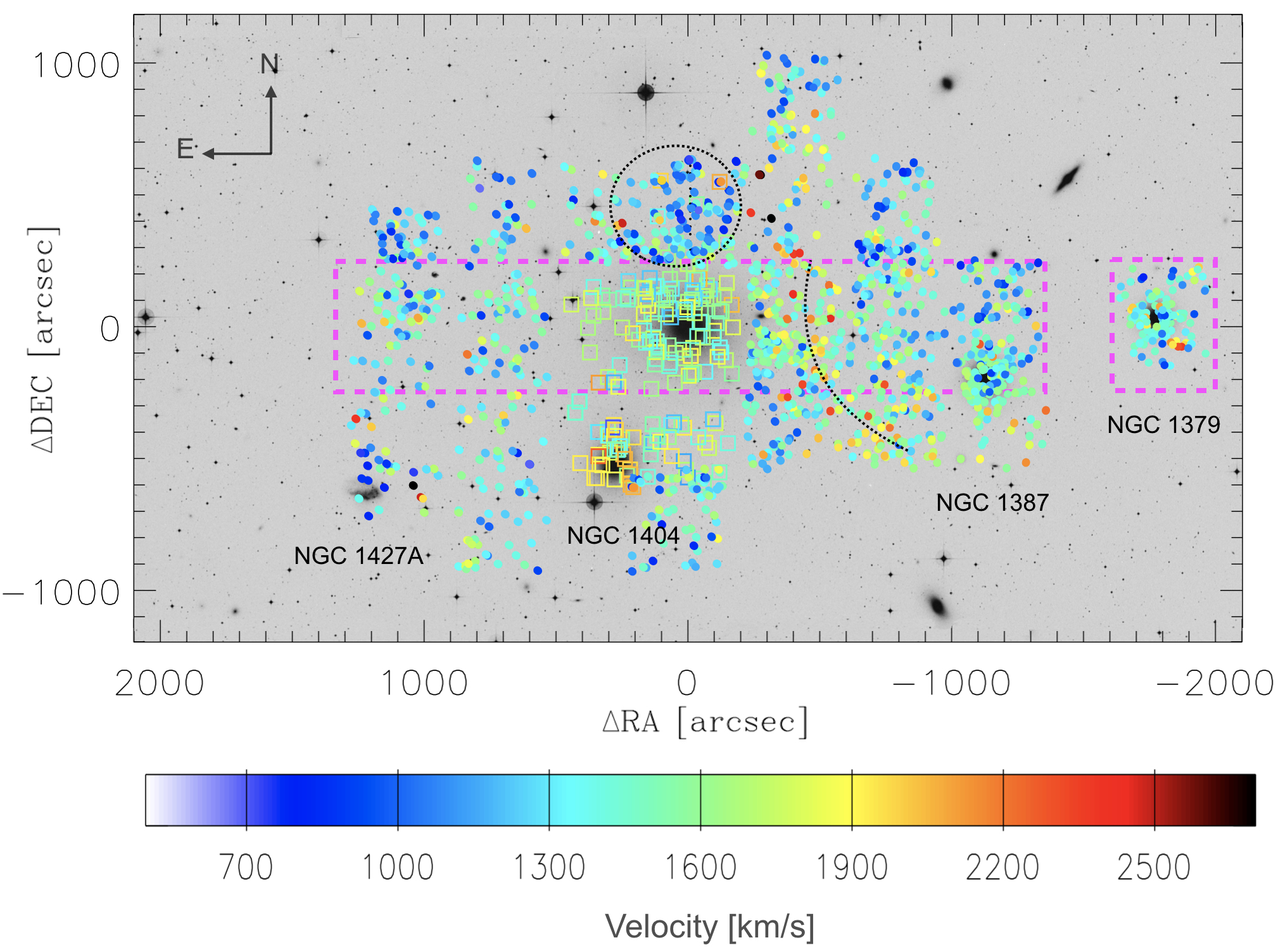}
 \caption{The final PNe catalog overplotted on a DSS image of the Fornax Cluster. Circles are objects identified in this work, squares are the PNe already presented in MN10. The PNe are color-coded by their velocity. Two streams of intracluster PNe can be highlighted from the image: a low-velocity PNe patch north of NGC~1399 (identified by a dotted big circle) and a high-velocity PNe bridge connecting the central galaxy with NGC~1387 (identified by the dotted curved line). See text for more detalis. The magenta slit overplotted on the PNe is the slit we define to obtain the velocity dispersion radial profile (corresponding to magenta points in Fig.~\ref{fig:pne_sigma}).}
 \label{fig:pne_velocity}
\end{figure*}
  
\section{Results}
\subsection{The final PN Velocity sample}
The final catalog of PNe with measured velocities comprises 1635 objects (of which 1452 new detections) and extends conspicuously both in number and spatial coverage all  previous overlapping PNe catalogs of the Fornax Cluster. 
The phase space distribution diagram is shown is Figure~\ref{fig:phasespace}. Planetary Nebula velocities are plotted as function of radius (calculated as distance from NGC~1399, in circular radii) whereas NGC~1404, NGC~1379, NGC~1387 and NGC~1427A are plotted as colored circles. 
NGC~1399 is plotted as black triangle and its systemic velocity is showed as a dotted black line through the figure. 
Note that the absence of data at radii $1450\arcsec<$ R $<1650\arcsec$ is due to the geometrical coverage of our pointings. 

Figure~\ref{fig:pne_velocity} shows the spatial location of the planetary nebulae identified in this work (circles) and those presented in MN10 (squares) color-coded by their velocity. 
The PNe are overplotted on a DSS image of $\sim 70 \times 40$ arcmin.  
An interesting result that one can infer from this Figure is the presence of streams that might be directly related to the history of the cluster as a whole, tracing recent streams falling into the cluster (highlighted in the figure with dotted lines).
Already MN10 reported the presence of a low-velocity PNe sub-population moving at about 700\kms slower than NGC~1399. We confirm this detection finding $\sim50$ PNe with 700\kms$<V<1000$\kms distributed between $+300\arcsec\le\Delta\delta\le +700\arcsec$ north of  NGC~1399 (dotted big circle in the figure). 
Thanks to our more extended coverage, we also highlight a high-velocity PNe stream extending for $\sim 700\arcsec$  in Dec to the west side of NGC~1399, connecting the central galaxy to NGC~1387, which confirms the findings of FDS-I. 
A more detailed analysis of PN velocities streams and a direct comparison with 2D kinematical maps obtained with GCs covering a similar region (presented in FVSS-I) and with deep photometry (from the FDS) will be presented in a forthcoming paper of the series.  
Here we wish to stress the similarity and good spatial agreement with FDS-I and \cite{Iodice17} that detected this previously unknown region of intracluster light (ICL) and also related it to an over-density in the population of blue globular clusters \citep{DAmbrusco16, Cantiello17}.  
These authors found that the ICL in this region contribute $\sim 5\%$ to the total light of the brightest cluster member. They also compared their findings to theoretical predictions for the ICL formation and support a scenario in which the intracluster population is built up by the tidal stripping of material from galaxy outskirts in close passages of the central galaxy. 

\begin{figure*}
 \includegraphics[width=17.5cm]{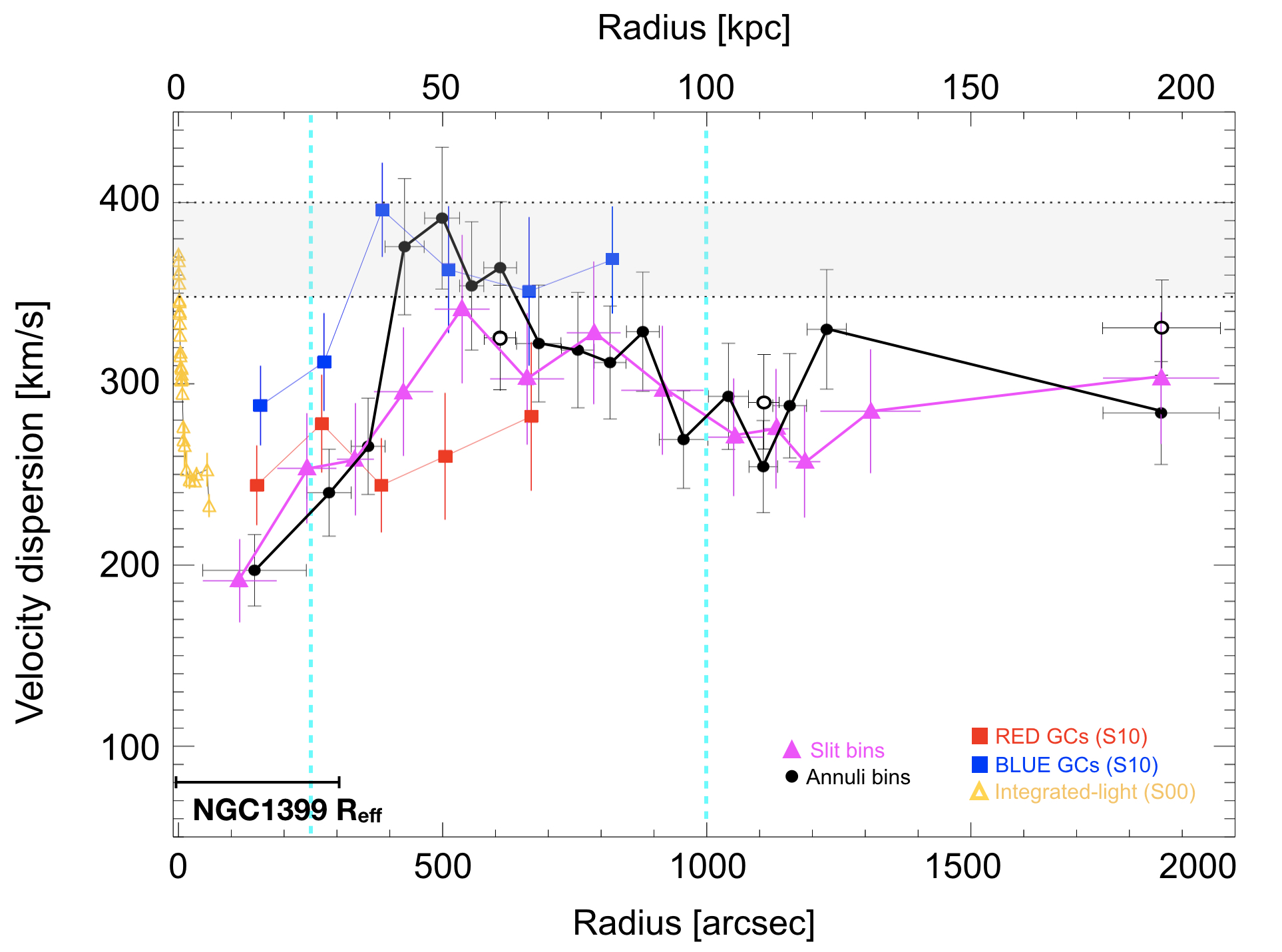}
 \caption{Line-of-sight velocity dispersion as a function of projected radius from the center of NGC~1399. The velocity dispersion obtained from our final catalog of PNe is plotted as filled black circles for the circular annuli calculation and as filled magenta triangles for the slit-bins. The three empty symbols correspond to the velocity dispersion inferred at a given radius removing the contribution of the PNe that are more likely associated to companion galaxies (all PNe within the effective radius of the corresponding galaxy). Red and blue globular clusters presented in Schuberth et al. (2010, S10 in the figure) are overplotted as red and blue squares, respectively, while  the integrated-stellar-light profile from Saglia et al. (2000, S00 in the figure) is showed as yellow triangles (major axis). Finally, the grey shaded horizontal region represents the velocity dispersion of the Fornax cluster with its one sigma error.}
 \label{fig:pne_sigma}
\end{figure*}

\subsection{The radial velocity dispersion profile}
Thanks to the large spatial coverage of our PNe catalog, we can trace the 
velocity dispersion of NGC~1399 out to large radii, well beyond the integrated-stellar-light measurements \citep{Saglia00}.  
This is crucial if one aims at covering the transition from the extended galaxy halo to the cluster potential \citep{Gerhard07,Dolag10}. 
To estimate the LOS velocity dispersion as a function of projected radius, we define circular apertures (annuli) at different radial distances from NGC~1399 and calculate the standard deviation of all the PNs in each of them. 
Given the irregular final spatial coverage of our sample, which extends more in RA than in Dec, we also follow a different approach and simulate a slit that extends  2600 arcsec in the EW direction ($\pm 1300\arcsec$ from the center of NGC~1399) and 600 arcsec in the NS direction ($\pm 300\arcsec$ from the center of NGC~1399). We choose this slit-width and orientation to maximize the number of PNe used and at the same time to have an homogeneous and complete spatial coverage. Moreover, we did not orient the slit along the major axis of NGC~1399 because we are not interested in features belonging to the central galaxies but we want to map the kinematics of the Fornax Cluster core. We then divide the slit in distance bins. 
The radial limits of both the circular annuli and the slit-bins are set such that all the bins have about the same number of  objects (90-91 in the first case, 69-70 in the second). 
Moreover, in both cases, we add an additional point considering the PNe around NGC~1379.  
A visualization of the slit and the additional point around NGC~1379 is shown in Figure~\ref{fig:pne_velocity} as magenta dashed boxes.
In Table~\ref{tab:one} we report the computed values for the velocity dispersion in all radial bins for the two approaches we undertook. 
Errors on the velocity dispersion values were estimated using Figure~7 in \cite{Napolitano01}, considering that we have 90 (70) PNe for each anulus (slit-bin).   
Finally, we added two additional data points (empty symbols) corresponding to the radii where there is an overlap with cluster galaxies in the field (namely NGC~1387 and NGC~1379) where we have recomputed the velocity dispersion excluding the PNe that in first approximation could be bound to the corresponding galaxies. We use the same 'spatial' criterion defined in Sec. 4.1 and used in Figure~\ref{fig:sys_vel}. We caution the reader that, however, this is only a first-order approximation and that a more robust classification will be performed in forthcoming papers.  

Figure~\ref{fig:pne_sigma} shows our LOS velocity dispersion profile (filled black points for the annuli and magenta points for the slit) compared to different collections of data from literature. 
In particular, we overplot the velocity distribution values for the red and blue globular clusters presented in \cite{Schuberth10} (red and blue squared, respectively)\footnote{We plot the full data set from \cite{Schuberth10} without interloper removal and including the GCs within 3 arcsec of NGC1404 and those with extreme velocities. We thrust this is an unbiased approach to carry out the comparison of GCs and PNe LOSVDs in the surveyed regions.} 
and the integrated-stellar-light profile along the major axis of NGC~1399 from \cite{Saglia00} (yellow triangles). Finally, the grey shaded horizontal region in the figure represents the $1\sigma$ contours of the velocity dispersion of the Fornax Cluster ($\sigma_{\rm cl} = 374\pm 26$ \kms) reported in \cite{Drinkwater01}, hereafter D01. We go twice as far from NGC~1399 compared to all the previously published velocity dispersion profiles, covering a radial distance of $\sim 2000\arcsec$ ($\sim 200$ kpc) from NGC~1399. 

In the plot, we identify three regions delimited by two characteristic radii, separated in the figure with dotted cyan lines:
\begin{itemize}
\item[1)] \emph{${\rm R}<250\arcsec$}. Here the PN velocity dispersions match the gradient found by \cite{Saglia00} from integrated-light and agree well with the dispersions reported in \cite{Schuberth10} for the red, metal-rich population of GCs that trace the halo of NGC~1399. PNe trace the kinematics of the central galaxy and in this region the population of {\sl virialized} stars dominates, based on results from cosmic simulations (e.g. \citealt{Cooper13, Cooper15}).  
Indeed \cite{Napolitano02} modeled the regions within $\sim400\arcsec$ and found dynamical signature at around $\sim200\arcsec$ of non equilibrium, consistent with the presence of a non mixed population possibly in relation with the interaction with the cluster potential or with companion galaxies (e.g. NGC~1404).
\item[2)] \emph{ $250\arcsec<{\rm R}<1000\arcsec$}. The PNe velocity dispersion rise steeply around 300-400 arcsec ($\sim1 -1.5$R$_{\rm eff}$) from  NGC~1399, reaching a peak around $\sim 500\arcsec$, similar to the blue, metal-poor GCs. 
As already stressed, a similar 'transition' region has been identified by FDS-I who also performed a cross-analysis joining light, GCs and X-Ray \citep{Paolillo02} and highlighting that all these tracers show a different slope with respect to the innermost regions (see Fig. 13 in FDS-I).   
In their Sec.~4.4, FDS-I highlight that the shape of the external light profile can be used to study the accretion mechanisms responsible for the mass assembly of galaxy clusters and that only kinematics can disentangle between "broken" light profiles which are associated to short and recent accretion of massive satellites \citep{Deason13} and "shallower profiles" (with excess of light) which indicate minor mergers events that took over long timescales where stars have undergone more mixing.  
In fact, in the first case (accretion of a massive satellite), a peak in velocity dispersion can be observed  for radii larger than the break radius. Adding the missing piece (kinematics information up to large radius) to the study of FDS-I, we show here that this is indeed the case for Fornax.  
In this region, of course, also PNe bound to NGC~1404 might contribute in the rise in velocity dispersion. However, in the slit approach (magenta points), where  NGC~1404 and the associated PNe are not present by construction, the rise is still present, although less pronounced.  
\item[3)] \emph{${\rm R } >1000\arcsec$}. Finally, at this distances our measurements flatten out, at a value of $\sigma_{\rm ICPNe} \sim 300$\kms, higher than the velocity dispersions measured for the single galaxies and closer (but slightly lower) to the value reported by D01 for the Fornax Cluster ($\sigma_{\rm cl}$). Here the PNe trace the cluster potential, measuring the kinematics of the intracluster light. 
We note also that our $\sigma_{\rm ICPNe}$ is in perfect agreement with the velocity dispersion of the giants galaxies in the Fornax Cluster ($308\pm30$\kms), that according to D01 are virialized\footnote{D01 find a difference of $\sim 130$\kms in velocity dispersion between the dwarf population and the giant population in the Fornax Cluster, which is consistent with the expected ratio of 2:1 for infalling and virialized, as predicted from \cite{Colless96}}. 
We also note that the value of the velocity dispersion that we infer when removing the PNe within 1 R$_{\rm eff}$ of NGC~1379 in the last bin is larger (330$\pm 38$\kms) and almost reach the $\sigma_{\rm cl}$ (empty black circle in Figure~\ref{fig:pne_sigma}). 
\end{itemize}

\begin{table}
\caption{Veolcity dispersion measurements for the two binning approaches.  By constructions, the bins have the same number of  objects (90-91 in the first case, 69-70 in the second) . }
\label{tab:one}
\begin{center}
\begin{tabular}{cc | cc}
\hline
Radius (annuli) & Vel. Disp. (annuli) & Radius (slit) & Vel. Disp. (slit) \\
arcsec & \kms  & arcsec & \kms  \\
\hline
     $144 \pm 98$ & $197 \pm 20$ & $116 \pm 70$ & $191  \pm 23$\\ 
     $285 \pm 42$ & $240 \pm 24$   & $243 \pm 56$ & $253 \pm 27$\\
     $359 \pm 32$ & $265 \pm 26$   & $335 \pm 35$ & $258 \pm 28$\\
     $428 \pm 37$ & $376 \pm 37$   & $426 \pm 56$ & $295 \pm 33$\\
     $499 \pm 33$ & $391  \pm 39$  & $537 \pm 52$ & $341 \pm 39$\\
     $555 \pm 23$ & $354 \pm 36$   & $660 \pm 70$ &  $303 \pm 38$\\
     $609 \pm 31$ & $364 \pm 36$   & $786 \pm 51$ &  $328 \pm 41$\\
     $682 \pm 42$ & $322 \pm 32$   & $916 \pm 78$ & $296 \pm 33$\\
     $756 \pm 31$ & $319 \pm 32$   & $1051 \pm 56$ &  $271 \pm  30$\\    
     $817 \pm 30$ & $312 \pm 31$   & $1131 \pm 24$ &   $275  \pm 30$\\
     $879 \pm 31$ &  $329 \pm 33$  & $1185 \pm 30$ & $257 \pm  28$\\
     $956 \pm 46$ & $269 \pm 27$   & $1310 \pm 95$ & $285 \pm   32$ \\
   $1041 \pm 38$ & $293 \pm 29$  &  $ 1860 \pm 83$ &  $303  \pm  38$\\
    $1107 \pm 27$ & $254 \pm 25$  &								&				\\ 
    $1157 \pm 32$ & $288 \pm 28$  &								&				\\
    $1227 \pm 37$ & $330 \pm 33$ &								&				\\
    $1860 \pm 150$ & $284 \pm 28$ &								&				\\
  \hline
\end{tabular}
 \begin{tablenotes}
  \item {\sl Col.~1, 3:} Distances from NGC~1399 of the annuli and slit bins, respectively. 
  \item {\sl Col.~2, 4:} Velocity dispersions calculated as standard deviation of all the PNe belonging to the correspondent bin for the circular annuli (col.2) and slit (col.4) approach. Uncertainties are assumed to be equal to 10\% of the measurement for the first case and 12\% of the measurement (extrapolated from Fig.~7 of \citealt{Napolitano01}, given the number of PNe in each bin).
  \end{tablenotes}
  \end{center}
\end{table}

\subsection{PNe as tracer of the ICL}
As a prelude to a more rigorous dynamical analysis, here we try to assess whether the difference in $\sigma$ is caused by the fact that the PNe and the cluster galaxies have different density profiles ($\alpha_{\rm PNe} = 3.0$ and $\alpha_{\rm cl} = 2.0$) but still live in the same potential.  
We use Eq.~2, 3 and 4 in \cite{Napolitano14}, where they use: 
$\alpha \equiv -d {\rm ln} j/ {\rm ln} r$, $\gamma \equiv -d {\rm ln} \sigma^{2}/ d{\rm ln} r$ and  
$\beta = 1 - \sigma^{2}_{\theta}/\sigma^{2}_{r}$ with $j$ being the deprojection of the surface brightness and $\sigma_{\theta}$ and  $\sigma_{r}$ 
the azimuthal and radial components of the velocity dispersion in spherical coordinates.  
Following their prescriptions, we assume that $\alpha$, $\beta$ and $\gamma$ are constant with radius and we use the projected dispersion written as in \cite{Dekel05}: 
\begin{equation}
\sigma_{p}(R) = A(\alpha, \gamma) B(\alpha, \gamma, \beta) V_{0}^{2} R^{-\gamma}, 
\end{equation}
where:
\begin{equation}
A(\alpha, \gamma) = \dfrac{1}{(\alpha + \gamma)}\dfrac{\Gamma[(\alpha + \gamma -1)/2]}{\Gamma[(\alpha + \gamma)/2]}\dfrac{\Gamma[\alpha/2]}{\Gamma[(\alpha -1)/2]}
\end{equation}
\noindent and :
 \begin{equation}
 B(\alpha, \gamma, \beta) = \dfrac {(\alpha + \gamma) - (\alpha + \gamma -1)\beta}{(\alpha + \gamma) -2\beta} 
 \end{equation}

\noindent Thus, at a given radius and under the working assumption that PNe and galaxies live in the same potential (i.e. $V_{0}$ is the same for both), the projected velocity dispersions of the two components depend only on the factors $A(\alpha, \gamma)$ and  $B(\alpha, \gamma, \beta)$, which are functions of the anisotropy parameter ($\beta$), the dispersion slope ($\gamma$) and the 3D density slope ($\alpha$).
At fixed radius R$=1200\arcsec$ , where the DM dominates, we assume that the intrinsic $\sigma$ profiles are both flat ($\gamma_{\rm cl} = \gamma_{\rm PNe} = 0$), 
we take an isothermal profile for the cluster ($\alpha_{\rm cl} =2$) and 
measure the stellar density slope from FDS-I ($\alpha_{\rm PNe} =3$).   
As for the anisotropy parameter, we consider the Fig.~2 in \cite{Mamon05} where they found that cosmological simulations suggest that moving toward the outskirts, the galaxies prefer a radial anisotropy ($\beta = 0.3$ at r$_{200}$).  
Under these hypothesis, we infer $\sigma_{\rm cl } = 1.3\times \sigma_{\rm PNe} = 380$\kms,  which is perfectly consistent with the value reported in D01 and indeed confirms that the PNe at that distance do not trace the spheroidal galaxy component but share the dynamics and the potential of the cluster galaxies.  
This demonstrates that indeed we are mapping the region of transition between the extended halo of NGC~1399 and the intracluster light, and we go well beyond it, clearly separating the two components. 
We will investigate this point further when performing a detailed dynamical modeling.  
In particular, we plan to clearly and more rigorously distinguish PNe that are bound to the halos of the different galaxies to the ones that are, instead, tracing the ICL (ICPNe) and directly compare our results to those obtained with GCs of FVSS-I. 
We will focus on the PNe and GCs streams, linking their velocity dispersions with possible mechanisms able to generate the ICL \citep{Murante07, Gerhard07, Rudick09, Cui14}.

\section{Summary and Conclusions}
We have presented  {\sl the largest and most extended planetary nebulae (PNe) kinematic catalog ever obtained for the Fornax cluster.}
We have obtained velocities of  $1635$ ($1452$ of which newly measured) PNe in a region of about $50\arcmin \times 30\arcmin$ in the core of the Fornax cluster using a counter-dispersed slitless spectroscopic technique with data from FORS2 on the VLT. 
The catalog also completes and spatially extends the sample of $\sim 180$ PNe in the halo of NGC~1399 and around NGC~1404 published by \cite{McNeil10}. 

In this paper we described our techniques and methods to identify and  select planetary nebulae directly from the calibrated and registered images and we calculated their velocities. 
We use the PNe line-of-sight (LOS) velocity and velocity dispersion distributions to trace the intracluster light (ICL) within 200 kpc of the Fornax Cluster core.

In particular: 
\begin{itemize}
\item  We presented the final LOS velocity distribution of the full catalog (this work + MN10) that comprises 1635 PNe. 
\item From the 2D spatial distribution of velocities we identified two streams of PNe. One low-velocity "group", north of NGC~1399, which was already observed in MN10, and a high-velocity PNe "bridge", connecting NGC~1399 and NGC~1387. Interestingly, a faint bridge in the same region was reported in \cite{Iodice17} from deep photometry and in \cite{DAmbrusco16} and \cite{Cantiello17} from an over-density in the GCs.  
From the FDS-I photometry results, we argue that this stream between NGC~1399 and NGC~1387 might indicate gravitational interaction that, however, did not happen in a very recent epoch. Indeed the bridge is rather faint ($\mu_{r} \sim 28 $--$ 29$ mag/arcsec$^{2}$) and above all does not look like a long tails but rather like a "diffuse light" (e.g. FDS-I). Different is, for example, the case of the Virgo cluster, where many bright and much more extended streams and tidal tails have been observed \citep{Mihos05, Janowiecki10, Capaccioli15, Longobardi15a}. 
\item We obtained the LOS velocity dispersion as a function of projected radius from the center of NGC~1399 and we compared it to other velocity dispersion profiles from different kinematical tracers. We reach a distance of $\sim 2000\arcsec$, corresponding to roughly 200 kpc, from NGC~1399, extending by far all the previous velocity dispersion profiles ever presented. 
\item In the overlap region, we found a good agreement with the LOS $\sigma$ profiles obtained from GCs in \cite{Schuberth10}, in particular the velocity dispersion of the PNe within R$<250\arcsec$ from NGC~1399 agree well with the dispersion of the red GC population, which shares the dynamical history of the central galaxy itself. For radii R$\ge 400\arcsec$, instead, the PN velocity dispersion better fits with that measured for the blue cluster population and it is closer to the value of the velocity dispersion of the main Fornax Cluster reported in D01 ($\sigma_{\rm cl}$), but it does not reach it. We noted that the measured $\sigma_{\rm PNe} = 300$\kms is perfectly consistent with the velocity dispersion measured by \cite{Drinkwater01} for the giants galaxies in the Fornax Cluster ($308\pm30$\kms), that are virialized. We therefore concluded that PNe at all radii probed in this study are virialized. 
\item We concluded, consequently, that after R$\sim400\arcsec$  the PNe are not bound to the NGC~1399 central halo. 
We also note that for intermediate radii ($400\arcsec<$R$<1000\arcsec$) the presence of sub-haloes causes a rise in the PNe velocity dispersion measurements. 
\item Finally, at very large radii (R$\ge 1000$\arcsec), where the velocity dispersion profile is roughly flat, we found that the difference in velocity dispersion between $\sigma_{\rm cl}$ and the $\sigma_{\rm ICPNe}$, where ICPNe have a $\sigma$ which is $\sim 80$\kms lower, is consistent with a dynamical scenario where the PNe share the same potential of the Cluster but have different density profile from the galaxies ($\alpha_{\rm PNe} = 3.0$ at R$=1200\arcsec$, assuming $\alpha_{\rm cl} = 2.0$).  
\end{itemize}

This is the second paper of Fornax Cluster VLT Spectroscopic Survey (FVSS), that aims at studying in depth the assembly history of one of the nearby dense cluster environments.  
With the biggest PNe catalog ever collected, spectra of globular clusters (GCs, Pota et al. 2018, Paper I, to be submitted) and ultra compact dwarfs (UCDs) with VIMOS@VLT,  spectroscopic data on the 3 large ETGs (NGC~1399, NGC~1404, NGC~1387) and 7 dwarf galaxies (NGC~1396, FCC188, FCC211, FCC215, FCC222, FCC223, FCC227, Spiniello et al. 2018) with MUSE@VLT we have {\sl the most complete and uniform collection of dynamical tracers to characterise the stellar population of the different systems and the DM profile deriving in this way the baryonic and dark mass distribution in the core of Fornax ($\sim 200$ kpc) with a precision never reached until now. }

\section*{Acknowledgments} 
We thank the referee for his/her very detailed and constructive comments which led to a significant improvement of this manuscript. We warmly thank Dr. Vincenzo Pota for his helpful comments. 
CS has received funding from the European Union's Horizon 2020 research and innovation programme under the Marie Sklodowska-Curie actions grant agreement n. 664931. 
NRN and EI acknowledge financial support from the European Union's Horizon
2020 research and innovation programme under the Marie Sklodowska-Curie
grant agreement n. 721463 to the SUNDIAL ITN network. 
NRN, EI and MP acknowledge the support of PRIN INAF 2014 "Fornax Cluster Imaging and Spectroscopic Deep Survey". 
CT is supported through an NWO-VICI grant (project number 639.043.308). 
We are grateful to Prof. Kenneth Freeman and Prof. Mike Bessell, for sharing spectra of the galaxy NGC~1387 that allowed us to further check our calibrations.   
This research is based on observations collected at the European Organisation for Astronomical Research in the Southern Hemisphere under ESO programme 096.B-0412(A). The paper has made use of the NASA/IPAC Extragalactic Database (NED) which is operated by the Jet Propulsion Laboratory, California Institute of Technology, under contract with the National Aeronautics and Space Administration.







\bsp	
\label{lastpage}
\end{document}